%% file: main.tex
\documentclass[conference]{IEEEtran}

\usepackage{algorithmic}
\usepackage{textcomp}
\usepackage{url}
\usepackage{verbatim}
\usepackage{enumitem}
\usepackage{xspace}
\usepackage{makecell}
\usepackage{multirow}
\usepackage{colortbl}
\usepackage{listings}
\usepackage{fancyvrb}
\usepackage{svg}
\usepackage{booktabs}
\usepackage{amsmath}
\usepackage{framed}
\usepackage[most]{tcolorbox}
\usepackage{xcolor}
\colorlet{shadecolor}{black!5}
\usepackage[hidelinks]{hyperref}
\definecolor{mygreen}{RGB}{220,255,220}
\definecolor{myred}{RGB}{255,220,220}
\definecolor{mycyan}{RGB}{220,255,255}
\definecolor{highlight}{RGB}{200,230,255}
\definecolor{worst}{RGB}{255,200,200}
\definecolor{nkupurple}{RGB}{113,26,95} 

\newtcolorbox{RQBox}{
    colback=gray!10,     
    colframe=black,    
    arc=5pt,             % 圆角半径
    boxrule=0.8pt,       % 边框粗细
    left=6pt, right=6pt, % 左右边距
    top=6pt, bottom=6pt, % 上下边距
    boxsep=0pt,          % 内容与边框间距
    before upper={\parindent15pt}, % 首行缩进
    fontupper=\normalsize,    % 字体大小
}

\newcommand{\toolname}{AdaptPrint\xspace}

\ifCLASSINFOpdf
\else
\fi
\begin{document}
%
% paper title
% Titles are generally capitalized except for words such as a, an, and, as,
% at, but, by, for, in, nor, of, on, or, the, to and up, which are usually
% not capitalized unless they are the first or last word of the title.
% Linebreaks \\ can be used within to get better formatting as desired.
% Do not put math or special symbols in the title.
\title{AdaptPrint: Response-Adaptive Fingerprinting of Black-Box LLM Services}

\author{Yilin Li\textsuperscript{1,2}, Yifei Zhang\textsuperscript{1,2}, Guozhu Meng\textsuperscript{1,2,\textsuperscript{*}} \\
\textsuperscript{1}State Key Laboratory of Cyberspace Security Defense, Institute of Information Engineering,\\ Chinese Academy of
Sciences, China \\
\textsuperscript{2}School of Cybersecurity, University of Chinese Academy of Sciences, China \\
$ \{ $liyilin2023, mengguozhu$ \}$@iie.ac.cn, zhangyifei26@mails.ucas.ac.cn
}

% conference papers do not typically use \thanks and this command
% is locked out in conference mode. If really needed, such as for
% the acknowledgment of grants, issue a \IEEEoverridecommandlockouts
% after \documentclass

% for over three affiliations, or if they all won't fit within the width
% of the page, use this alternative format:
% 
%\author{\IEEEauthorblockN{Michael Shell\IEEEauthorrefmark{1},
%Homer Simpson\IEEEauthorrefmark{2},
%James Kirk\IEEEauthorrefmark{3}, 
%Montgomery Scott\IEEEauthorrefmark{3} and
%Eldon Tyrell\IEEEauthorrefmark{4}}
%\IEEEauthorblockA{\IEEEauthorrefmark{1}School of Electrical and Computer Engineering\\
%Georgia Institute of Technology,
%Atlanta, Georgia 30332--0250\\ Email: see http://www.michaelshell.org/contact.html}
%\IEEEauthorblockA{\IEEEauthorrefmark{2}Twentieth Century Fox, Springfield, USA\\
%Email: homer@thesimpsons.com}
%\IEEEauthorblockA{\IEEEauthorrefmark{3}Starfleet Academy, San Francisco, California 96678-2391\\
%Telephone: (800) 555--1212, Fax: (888) 555--1212}
%\IEEEauthorblockA{\IEEEauthorrefmark{4}Tyrell Inc., 123 Replicant Street, Los Angeles, California 90210--4321}}

% make the title area
\maketitle
\renewcommand{\thefootnote}{}
\footnotetext[1]{*Corresponding author.}
% As a general rule, do not put math, special symbols or citations
% in the abstract
\begin{abstract}
Black-box LLM services have emerged as a practical deployment paradigm. Nevertheless, their opacity also hinders the systematic assessment of security risks and complicates copyright auditing for model owners.
Black-box LLM fingerprinting, which identifies the underlying LLM identity through query-response interactions, offers a promising way to bridge this gap.
Existing approaches typically collect responses from target LLM services using a fixed set of queries and perform poorly in the presence of realistic and complex configurations (e.g., system prompt and sampling settings). To overcome these limitations, we propose \toolname, a response-adaptive fingerprinting method for revealing hidden LLM identities in black-box LLM services. \toolname integrates three progressive response consistency probing strategies: Direct Probing, Continuation Probing, and Follow-up Probing. \toolname determines the final LLM identity by performing similarity matching among candidate LLMs. Experimental results show that \toolname  significantly outperforms state-of-the-art methods among 27 candidate models, achieving Top-1, Top-3, and Top-5 accuracies of 80.6\%, 90.3\%, and 92.1\%.
\toolname also demonstrates strong robustness across different defense strategies and decoding parameters.
\end{abstract}

% no keywords

% For peer review papers, you can put extra information on the cover
% page as needed:
% \ifCLASSOPTIONpeerreview
% \begin{center} \bfseries EDICS Category: 3-BBND \end{center}
% \fi
%
% For peerreview papers, this IEEEtran command inserts a page break and
% creates the second title. It will be ignored for other modes.
% \IEEEpeerreviewmaketitle

\input{sections/introduction}

\input{sections/background}

\input{sections/approach}

\input{sections/evaluation}

\input{sections/discussion}

\input{sections/related-work}

\input{sections/conclusion}

\newpage

\bibliographystyle{IEEEtran}
\bibliography{ref.bib}
\appendix
\input{sections/appendix.tex}

% that's all folks
\end{document}

%% file: sections/introduction.tex
\section{Introduction}
With the rapid advancement of large language models (LLMs), different LLMs have been widely adopted as the backend of various LLM services. Due to differences in model capability, cost, and task preference, LLMs are typically selected to match the requirements of specific service categories, such as knowledge-based question answering, software development, and real-time translation~\cite{ong2025routellm,jo2025sparellm}. Such services are commonly realized either through self-hosting models on cloud platforms or through wrapping responses obtained from upstream LLM providers into downstream applications~\cite{maslej2025artificial,forefrontAI,ai21platform,awsLLMrouting}. In these service scenarios, users interact with the model by requesting queries and receiving responses, without direct access to the model parameters or gradients. Among these LLM services, some providers disclose the underlying model identity, thereby allowing users to better assess model capabilities and ensuring greater transparency in service pricing~\cite{cai2025you, gptNLwhitebox, deepinfra2025whitebox}. We refer to them as \textit{white-box LLM services}, where the underlying LLM identity is explicitly visible to users. In contrast, other providers do not reveal the specific model identity due to flexible upgrades, reduced exposure to model-specific attacks, and the need for customization and service flexibility~\cite{lin2026behavioral, li2025thinktrap}. We refer to them as \textit{black-box LLM services}, where users can only access the service interface without knowing the exact underlying model.

In this paper, we primarily focus on the potential security issues associated with black-box LLM services. On the one hand, many existing attacks towards white-box LLM services adopt an adaptive attack paradigm, in which attackers make use of the disclosed LLM identity and tailor their attack strategies to exploit model-specific vulnerabilities~\cite{zheng2025jailbreaking,yang2025guiding}. These attacks have been demonstrated to pose substantial security risks to LLM services~\cite{andriushchenko2025jailbreaking,geisler2025reinforce}. However, concealing the underlying model identity does not necessarily prevent model-specific adaptive attacks. Attackers who infer the hidden LLM identity can similarly apply adaptive attacks to exploit model-specific vulnerabilities \cite{pasquini2025llmmap}. On the other hand, when the identity of the underlying LLM is not disclosed, a malicious service provider may violate the copyright or licensing agreements associated with the model~\cite{yang2025challenge}, such as by directly deploying a model whose license prohibits commercial use. Therefore, identifying the underlying LLM identity is essential for assessing the real-world risks of black-box LLM services and auditing potential copyright or licensing violations. This task can be formulated as the problem of identifying LLM fingerprints in LLM services. Several prior studies have attempted to extract LLM fingerprints from black-box LLM services~\cite{pasquini2025llmmap,shao2026reading,gao2024model}. However, these methods suffer from inherent limitations. 

First, existing approaches typically rely on a fixed set of queries to probe target LLM services. This static querying strategy lacks the ability to adapt to service-specific guardrails, such as safety filters or intent recognition mechanisms, which may alter responses and consequently degrade fingerprinting effectiveness~\cite{rebedea2023nemo,yang2026peering}. Second, these methods perform poorly in realistic and complex configurations (e.g., system prompt, sampling settings, and prompting strategies). When the fixed queries are applied under diverse and sophisticated service configurations, the resulting outputs often become unstable, causing the same underlying model to be erroneously identified as having different fingerprints.

To overcome the above limitations, we propose \toolname, a response-adaptive fingerprinting method for identifying hidden LLM identities in black-box LLM services. To address the first limitation, \toolname adaptively generates benign and domain-relevant queries for the target LLM service. To address the second limitation, \toolname introduces a response-consistency-based paradigm, in which fixed queries are no longer used as the starting point for LLM fingerprint analysis. Instead, fingerprint analysis is initiated from the responses produced by the target model itself and is subsequently conducted based on those responses. 

Specifically, \toolname first adaptively generates queries towards the domain of the target LLM service and collects the resulting responses. It then builds a candidate model pool and performs similarity-based comparison. For fingerprint identification, \toolname integrates three response consistency probing strategies: Direct Probing, Continuation Probing, and Follow-up Probing, and aggregates the information collected from these strategies to perform fingerprint matching. Experimental results demonstrate that, across 18 distinct system prompts, and 27 candidate models, \toolname achieves Top-1, Top-3, and Top-5 accuracies of 80.6\%, 90.3\%, and 92.1\%, respectively, improving over the baseline by 32.5, 21.3, and 15.2 percentage points. In addition, \toolname remains effective under common defense mechanisms, such as safety check and intent detection, and its reliability is further demonstrated across different levels of randomness in the decoding parameters. \toolname performs well in open-set verification, which allows the target to be rejected as belonging to none of the candidate models, demonstrating its applicability beyond closed-set classification.

\noindent\textbf{Contributions.} We make the following contributions.

\begin{itemize}[leftmargin=*]
    \item We present a new perspective on LLM fingerprinting through response consistency probing. Instead of relying on a fixed set of predefined queries, \toolname adaptively generates probes and uses the target model's responses as anchors for fingerprinting. 
    \item We propose \toolname, a response-adaptive fingerprinting method for revealing LLM identities in black-box services. \toolname integrates three progressive response-consistency probing strategies: Direct Probing, Continuation Probing, and Follow-up Probing. The three probing strategies enable progressive, multi-dimensional response matching for robust LLM fingerprints. 
    \item We perform extensive experimental evaluations under diverse system prompts, sampling parameters, prompting strategies and different LLMs. The results show that \toolname consistently outperforms state-of-the-art (SOTA) black-box fingerprinting methods and remains effective across a variety of scenarios, demonstrating its reliability and general applicability.

\end{itemize}

%% file: sections/background.tex
\section{Background and Motivation}
\label{sec:background}
\subsection{LLM Fingerprinting}
LLM fingerprint is inspired by the general notion of a fingerprint, which can be defined as a compact representation of an object's salient characteristics. In the context of LLMs, an LLM fingerprint refers to an identifiable signature that characterizes a specific model and can be used to distinguish it from others. Accordingly, LLM fingerprinting denotes the process of extracting or verifying such signatures for model identification, with the goal of determining the specific identity of an unknown LLM. LLM fingerprinting can be categorized into \emph{white-box} and \emph{black-box} settings. White-box LLM fingerprinting leverages internal information such as model weights, gradients, or activation states, whereas black-box LLM fingerprinting can only rely on query–response interactions with the target LLM.

A closely related technique is LLM watermarking\cite{zhang2024remark,giboulot2024watermax,pan2024markllm}. Although both fingerprinting and watermarking can be broadly viewed as approaches for model identification, they differ in their original mechanisms. LLM fingerprints are typically derived from the intrinsic characteristics of a model, such as its parameters, representations, or behavioral patterns. In contrast, LLM watermarks are usually intentionally embedded into model outputs or generation processes, often through modifications to the model weights~\cite{pang2024no}. 
\subsection{Threat Model}
\begin{figure}[t]
    \centering
    \includegraphics[width=1.0\linewidth]{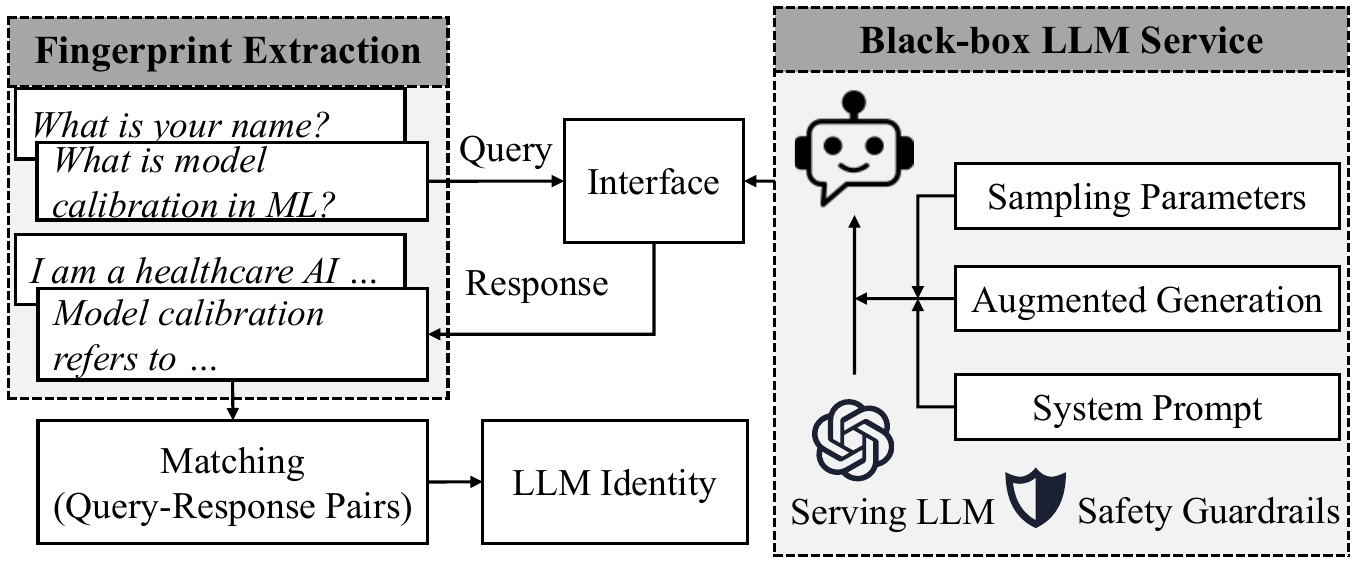}
    \caption{Threat model of black-box LLM fingerprinting.}
    \label{fig:treat_model}
\end{figure}

In this work, as shown in Figure \ref{fig:treat_model}, we concentrate on black-box LLM fingerprinting, as it aligns with the requirements of black-box LLM services. In practical black-box LLM services, providers commonly deploy either self-hosted LLMs or HTTP API-based LLMs as the underlying engine. From the perspective of users, interaction with such services is typically limited to a query–response interface (e.g., API or webpage), through which users submit queries and receive generated outputs. Access to internal model information, such as parameters, weights, or hidden representations, is generally unavailable. Black-box LLM services may also employ safety guardrails that filter malicious queries and responses. We assume that the LLM identity remains stable within the short query–response interaction window, as model updates or upgrades typically require longer deployment cycles~\cite{chauvin2026log}.

Formally, given an input and an LLM with specific configuration (system prompt, sampling parameters, prompting techniques, etc), the fingerprint extractor computes the fingerprint as:
\begin{equation}
fingerprint = \phi(\mathrm{llm}(\cdot, config), input),
\label{eq:fingerprint}
\end{equation}
where $\phi$ is the fingerprint extractor, $\mathrm{llm}(\cdot, config)$ denotes the target LLM under configuration config, and $input$ is the query used for fingerprinting.

\subsection{Motivation}
\begin{table*}[t]
\centering
\caption{Comparison with existing black-box LLM fingerprinting methods.}
\label{tab:previous-work}
\setlength{\tabcolsep}{7.5pt}
\begin{tabular}{ccccc} \toprule
     \textbf{Works} & \textbf{Question Source} & \textbf{Question Topic} & \textbf{Methodology} & \textbf{Underlying Assumption}  \\ \midrule
     MET~\cite{gao2024model} & Predefined & Wikipedia &  Maximum Mean Discrepancy  & $\phi(llm_{\mathrm{black}}, q) \approx \phi(llm_{\mathrm{white}}, q)$  \\
     LLMmap~\cite{pasquini2025llmmap}& Predefined & Safety Alignment  & Deep Learning  &  $\phi(llm_{\mathrm{black}}, q) \approx \phi(llm_{\mathrm{white}}, q)$  \\
     ZeroPrint~\cite{shao2026reading} & Predefined & Code & Jacobian Matrix & $\phi(llm_{\mathrm{black}}, q) \approx \phi(llm_{\mathrm{white}}, q)$   \\
     \toolname{} (ours) & Adaptive & Target's Topic &  Response Consistency Probing & $\phi(llm_{\mathrm{black}}, r^*) \approx \phi(llm_{\mathrm{white}}, r^*)$\\
     \bottomrule
\end{tabular}
\end{table*}

Table~\ref{tab:previous-work} summarizes prior work on black-box LLM fingerprinting. These studies offered valuable insights but still have limitations, which can be categorized into two aspects.

\noindent\textbf{Limitation 1.} Prior work commonly adopts predefined queries for interacting with LLM services, which may fail to accommodate diverse interaction contexts. For instance, the queries used in LLMmap~\cite{pasquini2025llmmap} include injection prompts, which may be detected by the safety guardrails of the target service and fail to collect the desired responses. Similarly, ZeroPrint~\cite{shao2026reading} focuses on code completion queries, when these queries are posed to LLM services designed for unrelated domains (e.g., medical assistant), they may activate the model's avoidance behavior or even trigger the service's intent-detection mechanism, resulting in a refusal to answer. \textit{This limitation highlights the need for designing benign and domain-relevant queries tailored to the application domain of the target LLM service.} Inspired by the Computerized Adaptive Testing (CAT) concept in psychology~\cite{wainer2000computerized,meijer1999computerized}, which dynamically selects test items based on a respondent’s answers during the assessment process, we aim to design an adaptive initial question generation method according to the domain of LLM service.

\noindent\textbf{Limitation 2.} Prior studies seek to develop LLM fingerprinting approaches that is insensitive to configuration variations. The underlying assumption of previous methods can be formalized as:
\begin{equation}
\phi(llm_{\mathrm{black}}, q) \approx \phi(llm_{\mathrm{white}}, q),
\end{equation}
where \(llm_{\mathrm{white}}\) denotes the known LLM that has the same underlying identity as the black-box LLM  \(llm_{\mathrm{black}}\) and $q$ denotes the input query. Therefore, they employ only a universal base configuration for fingerprint comparison. However, the assumption that configuration-dependent factors can be ignored becomes especially questionable when the system prompt and decoding parameters are complex and highly variable. For instance, when the system prompt is substantially longer than the user query, the generated response may be strongly influenced by information beyond the query itself. An analogy can be drawn from face recognition. A user's current appearance may deviate from their pre-enrolled profile due to makeup, hairstyle changes, or other uncontrolled factors. In such cases, an effective recognition method should adapt to the user's current appearance rather than relying exclusively on a fixed profile. Likewise, in LLM fingerprinting, the observable behavior of a target service can be affected by complex and variable configurations, including system prompts and decoding parameters. \textit{This leads to a key intuition: fingerprinting should be performed based on the LLM's actual responses under the current service setting, rather than relying on responses pre-collected under a presumed universal base configuration.} 

To demonstrate the validity of the above intuition, we conduct experiments on four representative models: Llama-3-8B-Instruct~\cite{meta_llama3_8b_instruct}, Llama-3.1-8B-Instruct~\cite{grattafiori2024llama}, Qwen2.5-7B-Instruct~\cite{yang2024qwen2}, and Qwen3-8B~\cite{yang2025qwen3}. We randomly selected a system prompt for docker-related tasks. For each model, we asked five different questions under temperature and top-p configurations sampled from [0.1, 0.3, 0.5, 0.7, 0.9] × [0.9, 1.0], resulting in 50 collected responses per model. We then sequentially compared the four models by computing the average negative log-likelihood (NLL) per token for each response, which provides a normalized measure of how well each response matches the generation distribution of a given model.
\begin{figure}[t]
    \centering
    \includegraphics[width=\linewidth]{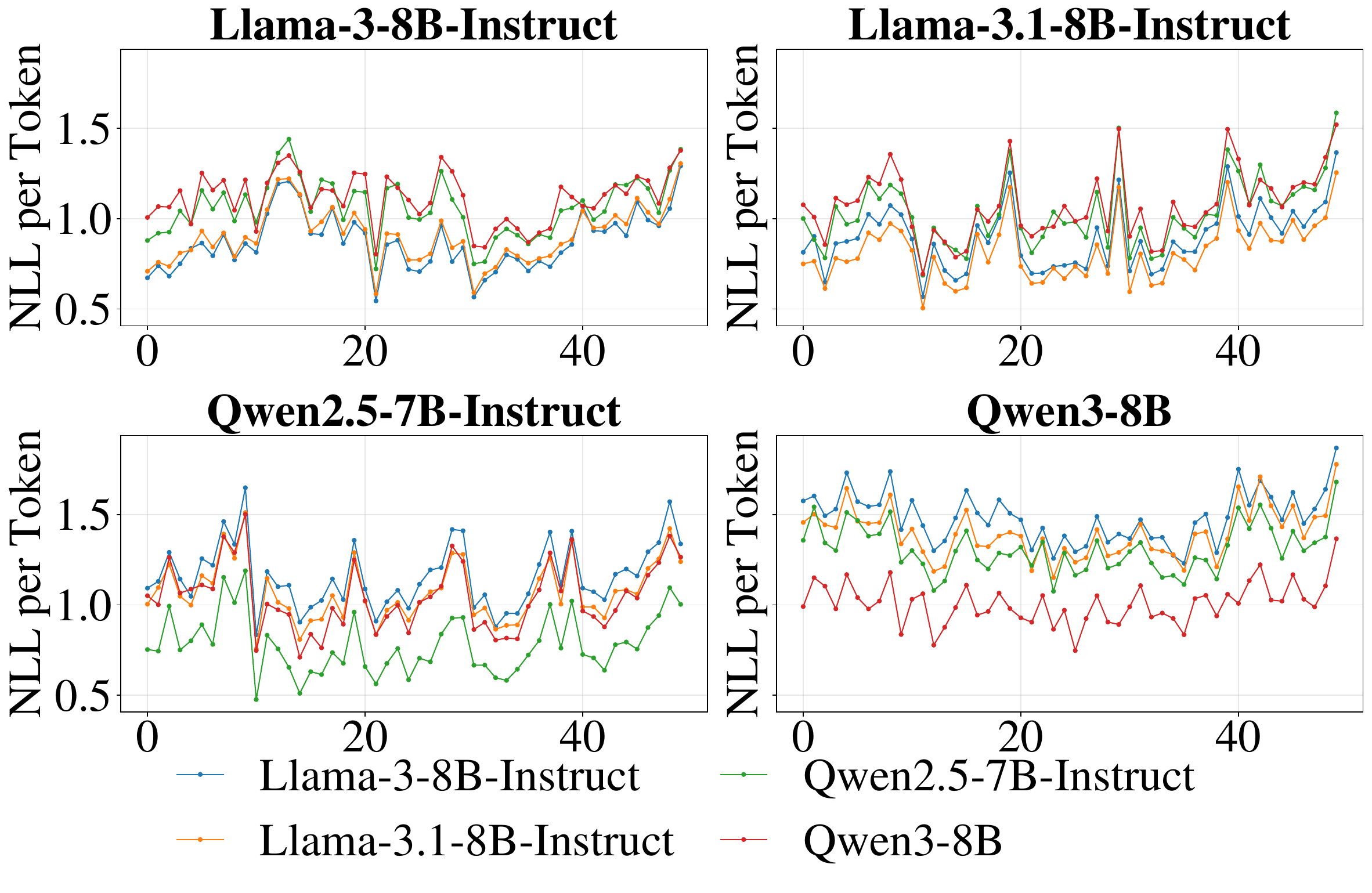}
    \caption{Self vs. other LLMs in NLL per token.}
    \label{fig:nll}
\end{figure}
The results are presented in Figure~\ref{fig:nll}. A lower NLL value suggests that the model assigns a higher likelihood to the corresponding response, indicating greater familiarity with it. The results show that each model generally yields the lowest NLL values for responses previously generated by itself. The result demonstrates the feasibility of developing LLM fingerprinting methods based on model responses. Based on the above analysis, the underlying assumption of \toolname can be formalized as:
\begin{equation}
r^* = llm_{\mathrm{black}}(q^*), 
\phi(llm_{\mathrm{black}}, r^*) \approx \phi(llm_{\mathrm{white}},r^*),
\end{equation}
where $r$ denotes the generated response. \toolname aims to construct a response-adaptive fingerprinting method while reducing the impact of complex and variable configuration settings in black-box LLM services.

%% file: sections/approach.tex
\section{Approach}
\begin{figure*}[t]
    \centering
    \includegraphics[width=0.95\linewidth]{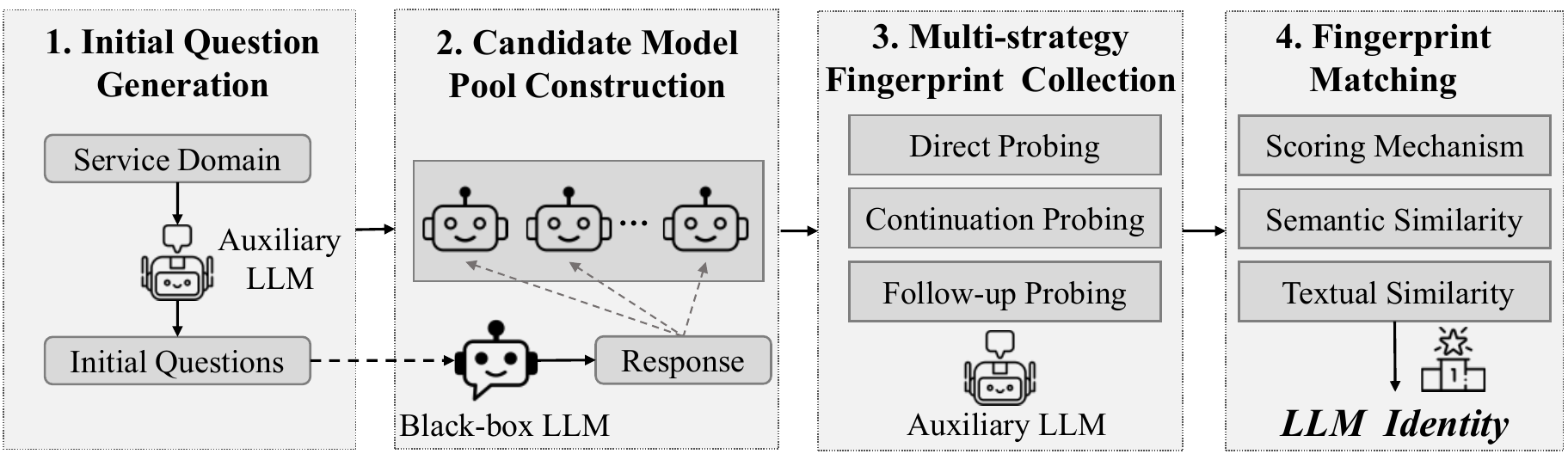}
    \caption{Workflow of \toolname{}.}
    \label{fig:overview}
\end{figure*}
As shown in Figure~\ref{fig:overview}, \toolname comprises four stages: initial question generation, candidate model pool construction, multi-strategy fingerprint collection, and fingerprint matching. First, \toolname generates domain-specific and low-entropy yet non-trivial questions tailored to the task domain of the target LLM service. Next, it constructs a pool of candidate models under a unified system prompt. \toolname then collects model fingerprints using three response consistency probing strategies: Direct Probing, Continuation Probing, and Follow-up Probing, which progressively probe the model behavior from different perspectives. Finally, \toolname performs similarity matching between the target model and candidate models across the three strategies, thereby determining the exact LLM identity.

\subsection{Theoretical Formulation}
The core objective addressed by \toolname is to identify the hidden LLM identity behind a black-box LLM service through query-response pairs.
Following the formulation in Section~\ref{sec:background}, we use \(llm_{\mathrm{black}}\) to denote the target black-box LLM service and
\(llm_{\mathrm{cand}}^{m}\) to denote a candidate LLM with identity \(m\). Among all candidates, \(m^{*}\) denotes the
ground-truth identity of \(llm_{\mathrm{black}}\).

Given an initial question \(q\), the target black-box LLM generates a response under its unknown deployment configuration:
\begin{equation}
r^{B} = llm_{\mathrm{black}}(q, config_{B}),
\end{equation}
where \(config_{B}\) includes factors such as the hidden system prompt, decoding parameters, and prompting techniques. A straightforward comparison is to query each candidate LLM with the same question \(q\). The underlying assumption is that the black-box LLM should behave most similarly to the candidate LLM with the same underlying identity:
\begin{equation}
\phi(llm_{\mathrm{black}}, q)
\approx
\phi(llm_{\mathrm{cand}}^{m^{*}}, q).
\end{equation}
However, this comparison can be fragile because the response of \(llm_{\mathrm{black}}\) may be strongly affected by its unknown service configuration.
% Instead of attempting to recover or approximate this hidden configuration, \toolname places all candidate LLMs under a unified reference configuration
% \(config_{0}\). This provides a common starting point for all candidate LLMs and avoids introducing additional prompt-specific bias into the candidate model pool. This design is based on the response-consistency property of LLM identities: the same underlying LLM should remain closer to itself than to different LLMs, even when observed under different
% configurations. Although \(config_{B}\) and \(config_{0}\) may differ, \(llm_{\mathrm{cand}}^{m^{*}}\) is expected to produce responses more consistent
% with the black-box responses than \(llm_{\mathrm{cand}}^{m}\), where \(m \neq m^{*}\).
The design of \toolname leverages response consistency: the target LLM's response is less sensitive to configuration changes than its query. Formally, let \(\mathcal{T}\) denote the set of probing strategies used by \toolname. For each strategy \(\tau \in \mathcal{T}\), \toolname constructs a response-conditioned probe:
\begin{equation}
q_{\tau} = P_{\tau}(q, r^{B}),
\end{equation}
where \(P_{\tau}(\cdot)\) denotes the probe construction function for strategy \(\tau\). The corresponding response of candidate model \(m\) is:
\begin{equation}
r^{C}_{m,\tau}
=
llm_{\mathrm{cand}}^{m}(q_{\tau}, config).
\end{equation}

The fingerprinting objective is to identify the candidate LLM whose responses are most consistent with the black-box responses across multiple probing
strategies:
\begin{equation}
\hat{m}
=
\arg\max_{m \in \mathcal{M}}
\operatorname{Agg}_{\tau \in \mathcal{T}}
\left[
\operatorname{Sim}
\left(
r^{B}_{\tau},
r^{C}_{m,\tau}
\right)
\right],
\label{eq:objective}
\end{equation}
where \(\mathcal{M}\) is the candidate model pool, \(r^{B}_{\tau}\) is the black-box response used as the reference for strategy \(\tau\), \(r^{C}_{m,\tau}\) is
the response generated by candidate model \(m\) under the same strategy,
and \(\operatorname{Sim}(\cdot,\cdot)\) measures response similarity.
\(\operatorname{Agg}(\cdot)\) denotes the strategy-level aggregation function that combines consistency evidence from different probing strategies into a
final matching score. A successful fingerprint matching is achieved when the candidate with the highest aggregated score corresponds to the hidden identity of
\(llm_{\mathrm{black}}\).

\subsection{Initial Question Generation}
During the initial question generation stage, \toolname constructs low-entropy yet non-trivial questions that are both natural and domain-relevant. This design is necessary for reliable fingerprint extraction. Open-ended queries, such as ``write a romantic poem.'' may produce substantially different outputs even from the same model, while trivial factual queries, such as ``Who won the Nobel Prize in Physics in 1921?'' often lead to highly similar responses
across different models. Neither provides stable and discriminative behavioral evidence.

To address this issue, \toolname uses an auxiliary LLM to generate initial questions according to the task domain of the black-box LLM service. The generated question is required to guide the answer toward a single stable explanatory backbone, such as a process, a constraint-resolution path, or a decision dependency path. For instance, a product-comparison question may require the model to jointly consider user preferences, operational constraints, and long-term usability. These requirements induce a coherent decision path in which the model must resolve a fixed set of tradeoffs, thereby reducing response variability and encouraging structurally consistent outputs. These constraints reduce response randomness while preserving sufficient model-specific semantic and lexical variation for fingerprinting. The complete prompt used for initial question generation is provided in Appendix~\ref{sec:initial}.

\subsection{Candidate Model Pool Construction}

LLM fingerprinting is essentially concerned with determining whether a black-box LLM service is powered by a model of interest. In practical auditing or
security analysis, this problem can be generalized to identifying whether the black-box LLM corresponds to one of several candidate models. Therefore, \toolname constructs a candidate model pool from LLMs with known identities and uses it as the comparison space for fingerprint matching.

After obtaining the initial query questions, \toolname queries the target black-box LLM service to collect its responses. \toolname places all candidate LLMs under a unified reference configuration. Specifically, all candidate LLMs are queried with the same neutral system prompt, ``You are a helpful assistant.'' and the same low-randomness decoding setting. This design provides a common starting point for candidate comparison. The purpose of candidate model pool construction is therefore not to approximate the hidden system prompt of the black-box service, but to establish a controlled candidate-side configuration. Under this unified configuration, response differences among candidate LLMs are more directly attributable to model-specific behavior. The low-randomness decoding setting further standardizes output stability and reduces interference from sampling noise. Once the candidate model pool has been constructed, \toolname performs multi-strategy fingerprint collection based on the black-box responses and the candidate model pool.

\subsection{Multi-strategy Fingerprint Collection}
At the core of \toolname is multi-strategy fingerprint collection, through which \toolname progressively gathers fingerprints using three response-adaptive consistency probing strategies: Direct Probing, Continuation Probing, and Follow-up Probing. The three probing strategies compare the black-box LLM with candidate models through direct response matching, response continuation, and context-aware follow-up questioning, respectively.

\noindent\textbf{Direct Probing Strategy.} The direct probing strategy queries each model in the candidate pool using the same set of questions generated during the initial question generation stage. For each probing question, the candidate models independently produce their responses, which are then compared with the corresponding response by the black-box LLM. By keeping the probing questions identical across models, this strategy provides a direct basis for evaluating response similarities under a controlled input condition. The rationale behind this strategy is to generate discriminative questions that elicit model-specific responses. By comparing the responses of the black-box service with those of candidate models, \toolname can assess their behavioral similarity and identify the most likely underlying model.

\noindent\textbf{Continuation Probing Strategy.} The continuation probing strategy guides the candidate model pool to continue generation from the midpoint of the black-box LLM's response, and compares the continuation with the second half of the original response. The rationale is that the second half of the target response can be regarded as the model's continuation of the first half, where the first half serves as the user prompt. Because model responses are typically much longer than the original questions, a sufficiently long and highly similar prompt can better constrain the generation context and reduce the influence of other factors, such as the system prompt. In addition, continuing from a prefix that resembles the model's own response approximately places the model back on the next-token-prediction trajectory of the original response, making the subsequent generation more likely to match the original continuation than those of other candidate models.
\begin{figure}[t]
\centering
\begin{tabular}{p{0.95\linewidth}}
\toprule
\textbf{Algorithm 1: Continuation Starting Point Selection} \\
\midrule
\begin{minipage}{0.95\linewidth}
\begin{algorithmic}[1]
% \STATE \textbf{Input:} Target response $r$, local open-source LLM $M_{\mathrm{loc}}$, exclusion ratio $\rho$, window size $w$
% \STATE \textbf{Output:} Continuation starting points $\mathcal{P}=\{p_1,p_2,p_3\}$

% \STATE $\mathbf{t}=(t_1,t_2,\ldots,t_N) \leftarrow \operatorname{Tokenize}(r)$
% \STATE $\mathbf{z}_i \leftarrow M_{\mathrm{loc}}(\cdot \mid t_{<i}), \quad i=1,\ldots,N$
\item[\textbf{Input:}] Target response $r$, local open-source LLM $M_{\mathrm{loc}}$, exclusion ratio $\rho$, window size $w$
\item[\textbf{Output:}] Continuation starting points $\mathcal{P}=\{p_1,p_2,p_3\}$

\STATE $\mathbf{t}=(t_1,t_2,\ldots,t_N) \leftarrow \operatorname{Tokenize}(r)$
\STATE $\mathbf{z}_i \leftarrow M_{\mathrm{loc}}(\cdot \mid t_{<i}), \quad i=1,\ldots,N$
\FOR{$i = 1$ to $N$}
    \STATE Let $\ell_i^{(1)}$ and $\ell_i^{(2)}$ denote the Top-1 and Top-2 logits in $\mathbf{z}_i$
    \STATE $c_i \leftarrow \ell_i^{(1)} \cdot \left(\ell_i^{(1)}-\ell_i^{(2)}\right)$
\ENDFOR

\STATE $\mathcal{I} \leftarrow
\left\{
i \mid \lfloor \rho N \rfloor < i \leq N-\lfloor \rho N \rfloor
\right\}$

\FOR{each $j$ such that $\{j,\ldots,j+w-1\}\subseteq \mathcal{I}$}
    \STATE $a_j \leftarrow \frac{1}{w}\sum_{i=j}^{j+w-1} c_i$
\ENDFOR

\STATE $\mathcal{R} \leftarrow \operatorname{RankDesc}(\{j\}, a_j)$

\STATE $\mathcal{C}_1,\mathcal{C}_2,\mathcal{C}_3
\leftarrow \operatorname{KMeans}(\mathcal{R}, K=3)$

\FOR{$q = 1$ to $3$}
    \STATE $i_q \leftarrow
    \arg\min_{j \in \mathcal{C}_q}
    \operatorname{rank}_{\mathcal{R}}(j)$
    \STATE $p_q \leftarrow \operatorname{TextOffset}(i_q, r)$
\ENDFOR

\STATE \textbf{return} $\mathcal{P}=\{p_1,p_2,p_3\}$
\end{algorithmic}
\end{minipage}
\\
\bottomrule
\end{tabular}
\label{fig:continuation_start_selection}
\end{figure}
The continuation starting point should be selected at a position with low next-token entropy. Such positions provide a more aligned generation starting state across different models, thereby improving the comparability of their continuations. 

The detailed selection procedure is presented in Algorithm 1. \toolname first evaluates each token position using a heuristic selection score computed from a locally deployed LLM. The score jointly considers the top-1 logit and its margin over the top-2 logit. After excluding boundary tokens, \toolname aggregates confidence scores with a sliding window and ranks candidate positions accordingly. To obtain diverse starting points, the ranked positions are clustered into three groups, and the highest-ranked position in each group is selected as a continuation point. This design promotes stable continuations through high-confidence starting points, while selecting from three distinct groups introduces diversity and captures complementary information across continuations. We do not claim that these three points are globally optimal. Instead, this design serves as a practical and general solution, as the optimal continuation points are difficult to determine in a black-box setting and may differ across candidate models due to their distinct generation characteristics. Once the continuation starting point is determined, \toolname employs a standardized prompt to regulate the continuation behavior and collect responses from the candidate model pool. The complete continuing prompt is provided in Appendix~\ref{sec:continuing}.

\noindent\textbf{Follow-up Probing Strategy.} The follow-up probing strategy formulates an additional question conditioned on the black-box LLM's response, and provides the initial question-response context to both the black-box LLM and the candidate model pool when comparing their responses. Compared with the continuation strategy, this strategy introduces a longer contextual prefix into the user prompt. When this prefix is generated by the model itself, the attention-based probability distribution over the preceding context is expected to be closer to that of the black-box LLM's generation process, thereby increasing the likelihood of producing subsequent responses similar to those of the target black-box model. For the follow-up question generation, \toolname feeds the previous response into an auxiliary LLM and prompts it to generate a contextually coherent follow-up question based on the preceding response. The follow-up question maintains contextual continuity while extending the conversation. The full prompt is provided in Appendix~\ref{sec:following}.
\subsection{Fingerprint Matching}
In the fingerprint matching stage, \toolname performs the final fingerprint identification. 
Specifically, \toolname integrates the analysis results obtained from the three strategies above and compares the similarity between each LLM in the candidate model pool and the black-box LLM. 
\toolname measures response similarity from two complementary perspectives: semantic similarity and textual similarity.

For semantic similarity, \toolname uses a semantic embedding model to encode model responses into vector representations, and then compute their cosine similarity. 
This embedding-based comparison has been widely adopted in prior work as an effective way to measure semantic closeness between generated texts~\cite{behnamghader2024llm2vec,zhang2025qwen3,chen2024m3}. 
For textual similarity, \toolname adopts n-gram-based Jaccard similarity, which captures local contiguous patterns in text. Such patterns often reflect the model's lexical choices, phrase-level preferences, and local textual pattern~\cite{sari2018topic}. Given two responses $x$ and $y$, we denote their $n$-gram sets as $G_n(x)$ and $G_n(y)$, respectively. 
The n-gram-based Jaccard similarity is defined as:
\begin{equation}
\operatorname{Jaccard}_n(x,y)
=
\frac{|G_n(x)\cap G_n(y)|}{|G_n(x)\cup G_n(y)|}.
\end{equation}

\toolname uses a score-based mechanism to infer the most likely identity of the black-box LLM service. The detailed procedure is shown in Algorithm 2. \toolname aggregates evidence from multiple fingerprinting strategies. For each strategy, it computes a similarity score between the black-box response and each candidate model's response using both semantic and textual similarities. Specifically, semantic similarity captures high-level meaning consistency, while textual similarity captures local lexical and phrase-level overlap. For Continuation Probing, three generation branches are evaluated, and the two highest scores are retained and summed. For Direct Probing and Follow-up Probing, the strategy-level score is directly added to the candidate's total score.

\toolname adopts this aggregation rule to reflect the different sources of probing strategies. Continuation Probing is inherently sensitive to the choice of continuation starting point, as different positions may expose different generation behaviors. We therefore evaluate three diverse starting points to obtain complementary evidence, while retaining the top two scores to reduce the influence of an uninformative or unstable branch. In contrast, Direct Probing and Follow-up Probing operate on complete probing questions and each produces a single response comparison, so no additional branch aggregation is required. The final score of each candidate is aggregated across the three probing strategies, and the model with the highest total score is the inferred identity.

  \begin{figure}[t]
  \centering
  \begin{tabular}{p{0.95\linewidth}}
  \toprule
  \textbf{Algorithm 2: Score-based Fingerprint Matching} \\
  \midrule
  \begin{minipage}{0.95\linewidth}
  \begin{algorithmic}[1]
  \item[\textbf{Input:}] Black-box responses $\mathcal{Y}^{B}$; candidate responses $\mathcal{Y}^{C}$; candidate model set $\mathcal{M}$; similarity
  weights $w_1,w_2$; strategy weights $\lambda_1,\lambda_2,\lambda_3$
  \item[\textbf{Output:}] Inferred model identity $\hat{m}$

  \STATE $V(m) \leftarrow 0,\quad \forall m \in \mathcal{M}$
  \STATE $\mathcal{S} \leftarrow \{\textsc{Direct}, \textsc{Continuation}, \textsc{Follow-Up}\}$

  \FOR{each $s \in \mathcal{S}$}
      \IF{$s=\textsc{Continuation}$}
          \FOR{each $g \in \{1,2,3\}$}
              \FOR{each $m \in \mathcal{M}$}
                  \STATE $A_g(m) \leftarrow
                  w_1 \operatorname{Sim}_{\mathrm{sem}}(y^{B}_{s,g}, y^{C}_{m,s,g})
                  +
                  w_2 \operatorname{Sim}_{\mathrm{text}}(y^{B}_{s,g}, y^{C}_{m,s,g})$
              \ENDFOR
          \ENDFOR

          \FOR{each $m \in \mathcal{M}$}
              \STATE Select the two highest scores from $\{A_1(m), A_2(m), A_3(m)\}$ as $\mathcal{G}(m)$
              \STATE $V(m) \leftarrow V(m)
              +
              \lambda_2
              \sum_{a \in \mathcal{G}(m)} a$
          \ENDFOR

      \ELSE
          \FOR{each $m \in \mathcal{M}$}
              \STATE $A(m) \leftarrow
              w_1 \operatorname{Sim}_{\mathrm{sem}}(y^{B}_{s}, y^{C}_{m,s})
              +
              w_2 \operatorname{Sim}_{\mathrm{text}}(y^{B}_{s}, y^{C}_{m,s})$
          \ENDFOR

          \FOR{each $m \in \mathcal{M}$}
              \IF{$s=\textsc{Direct}$}
                  \STATE $V(m) \leftarrow V(m) + \lambda_1 A(m)$
              \ELSE
                  \STATE $V(m) \leftarrow V(m) + \lambda_3 A(m)$
              \ENDIF
          \ENDFOR
      \ENDIF
  \ENDFOR

  \STATE \textbf{return} $\hat{m}=\arg\max_{m \in \mathcal{M}} V(m)$
  \end{algorithmic}
  \end{minipage}
  \\
  \bottomrule
  \end{tabular}
  \label{fig:score_matching}
  \end{figure}

%% file: sections/evaluation.tex
\section{Evaluation}
To systematically evaluate \toolname, we conduct the experiments to answer the following research questions:

\begin{enumerate}[leftmargin=*,label=\textbf{RQ\arabic*.}]
\item \textbf{(Effectiveness and Robustness)} How does \toolname perform in identifying black-box LLMs across different models, LLM service settings, and security defense strategies?

\item \textbf{(Ablation Study)} How does each component of \toolname contribute to its overall effectiveness?

\item \textbf{(Applications)} Does \toolname remain effective in downstream tasks and real-world scenarios?
\end{enumerate}

\subsection{Evaluation Setup}
\noindent\textbf{LLM Selection. }We conduct experiments on a diverse set of model instances, spanning mainstream open-source and closed-source model families and multiple variants within the same family. The full list of evaluated models is reported in Table~\ref{tab:candidate_models}. To control the experimental cost, we select 12 representative models (marked with gray cells) as black-box targets and use all models as the candidate model pool. The target models are chosen to cover different model families and publishers, including Amazon, ByteDance, DeepSeek, Google, Inception, InclusionAI, Meta-Llama, MistralAI, OpenAI, Qwen, Stepfun, and Xiaomi. For each covered family, we select one representative model as the black-box target, while retaining all 27 models as candidates for fingerprint matching.

\noindent\textbf{LLM Configuration. }For decoding settings, we use six temperature/top-p pairs: \((0.0,1.0)\), \((0.1,0.9)\), \((0.3,0.9)\), \((0.5,0.9)\), \((0.7,0.95)\), and \((0.9,1.0)\). We choose \((0.7, 0.95)\) as the default temperature/top-p setting, and use the other combinations to evaluate the randomness of decoding parameters on the effectiveness of \toolname. These settings define a sampling spectrum from conservative to open-ended generation. For prompt configurations, we collect system prompts following an LLM-service style from public GitHub repositories~\cite{linexjlin2026gpts,0xeb2026thebigpromptlibrary,louisshark2026chatgptsystemprompt}, and further include a random subset of prompt configurations (including Chain-of-Thought and Retrieval-Augmented Generation template) from prior work~\cite{pasquini2025llmmap} for comparison. The information of the evaluated prompts is reported in Table~\ref{tab:prompt_configs}.

\noindent\textbf{Implement and Hyperparameters Settings. }We use OpenRouter~\cite{openrouter2026} as the unified model routing platform to query different LLMs. We conduct analysis that uses local LLMs on a server equipped with an Intel Xeon Gold 5218R CPU and four NVIDIA A800 GPUs. By default, we use Qwen-3-8B~\cite{yang2025qwen3} for continuation starting point selection and Qwen3-4B-Embedding~\cite{zhang2025qwen3} as the semantic embedding model. We use Gemini-3-Flash-Preview~\cite{google_gemini3_flash_preview} as the auxiliary LLM to perform tasks such as initial question generation. For the candidate model pool, we use a temperature/top-p setting of \((0.0,1.0)\). For Algorithm 1, we set the exclusion ratio to 0.25, the window size to 10. For Algorithm 2, \(w1\) and \(w2\) are 0.5, \(\lambda_1\), \(\lambda_2\), and \(\lambda_3\) are all set to 0.25.

\subsection{RQ1: Effectiveness and Robustness}

\begin{table*}[t]
\centering
\caption{Candidate LLMs used in our evaluation.}
\label{tab:candidate_models}
\setlength{\tabcolsep}{8.8pt}
\begin{tabular}{c l c l c l}
\toprule
\textbf{No.} & \textbf{Model Name} & \textbf{No.} & \textbf{Model Name} & \textbf{No.} & \textbf{Model Name} \\
\midrule
\cellcolor{gray!15}1  & \cellcolor{gray!15}deepseek/deepseek-v4-flash & 10 & openai/gpt-4.1-nano & 19 & meta-llama/llama-3.3-70b-instruct \\
\cellcolor{gray!15}2  & \cellcolor{gray!15}amazon/nova-micro-v1 & 11 & openai/gpt-4o-mini & 20 & meta-llama/llama-4-scout \\
\cellcolor{gray!15}3  & \cellcolor{gray!15}bytedance-seed/seed-1.6-flash & 12 & openai/gpt-oss-120b & \cellcolor{gray!15}21 & \cellcolor{gray!15}meta-llama/llama-4-maverick \\
\cellcolor{gray!15}4  & \cellcolor{gray!15}inclusionai/ling-2.6-flash & \cellcolor{gray!15}13 & \cellcolor{gray!15}google/gemma-4-31b-it & \cellcolor{gray!15}22
& \cellcolor{gray!15}mistralai/mistral-small-3.2-24b-instruct \\
\cellcolor{gray!15}5  & \cellcolor{gray!15}xiaomi/mimo-v2.5 & 14 & google/gemma-4-26b-a4b-it & 23 & mistralai/mistral-small-2603 \\
\cellcolor{gray!15}6  & \cellcolor{gray!15}stepfun/step-3.7-flash & 15 & google/gemma-3-27b-it & 24 & mistralai/mistral-large-2512 \\
\cellcolor{gray!15}7  & \cellcolor{gray!15}inception/mercury-2 & 16 & google/gemini-2.5-flash-lite & \cellcolor{gray!15}25 & \cellcolor{gray!15}qwen/qwen3-30b-a3b-instruct-2507 \\
8  & openai/gpt-3.5-turbo & 17 & google/gemini-3.1-flash-lite & 26 & qwen/qwen-2.5-7b-instruct \\
\cellcolor{gray!15}9  & \cellcolor{gray!15}openai/gpt-4.1-mini & 18 & meta-llama/llama-3.1-8b-instruct & 27 & qwen/qwen3-235b-a22b-2507 \\
\bottomrule
\end{tabular}
\end{table*}

\begin{table}[t]
\setlength{\tabcolsep}{4.5pt}
\centering
\caption{Overview of prompt configurations.}
\label{tab:prompt_configs}
\begin{tabular}{ccccc}
\toprule
\textbf{No.} & \textbf{Assistant Domain} & \textbf{Length} & \textbf{Type} & \textbf{Source}\\
\midrule
1 & Coding Assistance I & 2218 & Service Style & GitHub\\
2 & Coding Assistance II & 1474 & Service Style & GitHub\\
3 & Cold Email & 2571 &  Service Style & GitHub\\
4 & Database and SQL & 452 & Service Style & GitHub \\
5 & Docker Assistance & 2731 & Service Style & GitHub \\
6 & Event Planning & 1528 & Service Style & GitHub \\
7 & General Assistance& 1322 & Service Style & GitHub \\
8 & Medical Diagnosis I & 2325 & Service Style & GitHub \\
9 & OpenAI API & 1285 & Service Style & GitHub\\
10 & Shopping Assistance& 967 & Service Style & GitHub \\
11 & Founder Mentorship & 129 & Basic & LLMmap\\
12 & Museum Guide & 196 & Basic & LLMmap \\
13 & Pet Care & 261 & Basic & LLMmap \\
14 & Fitness Tracker& 193 & COT & LLMmap\\
15 & Travel Planning & 213 & COT & LLMmap \\
16 & Tutoring Assistance & 190 & COT & LLMmap\\
17 & Coding Troubleshooting & 266 & RAG & Self-Built\\
18 & Medical Diagnosis II & 262 & RAG & Self-Built\\
\bottomrule
\end{tabular}
\end{table}

To answer RQ1, we compare \toolname with the SOTA black-box LLM fingerprinting methods across diverse LLM service settings and under deployed defense strategies. We also evaluate the robustness of \toolname across decoding settings with varying levels of randomness and analyze its performance in open-set settings.

\noindent\textbf{Comparison with Baseline Methods.} We compare \toolname against three state-of-the-art black-box LLM fingerprinting baselines: MET~\cite{gao2024model}, LLMmap~\cite{pasquini2025llmmap}, and ZeroPrint~\cite{shao2026reading}. The details of their methods are shown in Table~\ref{tab:previous-work}. The experiments are conducted on the 18 different system prompts listed in Table~\ref{tab:prompt_configs}, with the target black-box LLM using a temperature/top-p setting of \((0.7,0.95)\).

We report Top-1, Top-3, and Top-5 accuracy to evaluate whether the correct model can be identified among the candidate models. The baseline methods rely on template-library-based fingerprint matching. Specifically, they first extract fingerprints from candidate LLMs and then compare the black-box model against these precomputed fingerprints. To construct the reference fingerprint library, we consider two variants: \textbf{Method-Gen} and \textbf{Method-All}. Method-Gen uses fingerprints extracted with a general system prompt as the reference library. Method-All uses all instances from the other system prompts to construct the reference library. To align with the setting of Method-All, the candidate models of \toolname are also restricted to the same 12 different target models. We evaluate the performance of \toolname{} using three different initial questions. \toolname{}-Dual denotes the average result obtained from two different initial questions, while \toolname{}-Triple denotes the result obtained from all three initial questions.

\begin{table*}[t]
\centering
\caption{Comparison of different methods.}
\label{tab:main_experiment}
\setlength{\tabcolsep}{1.8pt}
% \resizebox{\textwidth}{!}{
% \begin{tabular}{lcccccccccc}
% \begin{tabular}{l*{8}{>{\centering\arraybackslash}p{1.8cm}}}
\begin{tabular}{
    *{4}{>{\raggedright\arraybackslash}p{1.6cm}}
    *{4}{>{\raggedright\arraybackslash}p{1.9cm}}
    >{\raggedright\arraybackslash}p{2.4cm}
}
\toprule
\textbf{Metric} 
& \textbf{LLMmap}  
& \textbf{Met} 
& \textbf{ZeroPrint} 
& \textbf{\toolname-Q1}
& \textbf{\toolname-Q2}
& \textbf{\toolname-Q3}
& \textbf{\toolname-Dual-Avg}
& \textbf{\toolname-Triple}
\\
\midrule
\textbf{Top-1-gen} & 36.6\% & 7.5\% & 48.1\% & 56.0\% & 63.0\% & 65.7\% & 75.5\% & \textbf{80.6\%(+32.5pp)} \\
\textbf{Top-3-gen} & 63.9\% & 21.0\% & 69.0\% & 82.4\% & 83.3\% & 81.9\% & 89.0\% & \textbf{90.3\%(+21.3pp)} \\
\textbf{Top-5-gen} & 74.1\% & 31.6\% & 76.9\% & 88.4\% & 89.4\% & 86.6\% & 91.5\% & \textbf{92.1\%(+15.2pp)} \\
\textbf{Top-1-all} & 57.4\% & 4.2\% & 78.2\% & 72.2\% & 78.7\% & 75.0\% & 85.6\% & \textbf{87.0\%(+8.8pp)} \\
\textbf{Top-3-all} & 73.1\% & 7.1\% & 84.7\% & 88.4\% & 90.3\% & 87.0\% & 91.4\% & \textbf{91.7\%(+7.0pp)} \\
\textbf{Top-5-all} & 81.5\% & 9.9\% & 88.9\% & 94.4\% & 93.5\% & 92.6\% & 94.3\% & \textbf{94.4\%(+5.5pp)} \\
\bottomrule
\end{tabular}
\end{table*}

Table~\ref{tab:main_experiment} reports the experimental results. Met yields the lowest performance among all evaluated methods. This is mainly due to its limited ability to distinguish between different LLMs: many models are often assigned the same score, which reduces its effectiveness for fine-grained identification. LLMmap is effective to some extent, but its performance on complex system prompts can still be improved. Zeroprint achieves strong performance, with Top-1, Top-3, and Top-5 accuracies of 48.1\%, 69.0\%, and 76.9\% on Method-Gen and 78.2\%, 84.7\%, and 88.9\% on Method-All, respectively. Nevertheless, \toolname consistently outperforms the baselines, improving Top-1 accuracy by 32.5 and 8.8 percentage points on Method-Gen and Method-All, respectively.

\begin{figure*}[t]
    \centering
    \includegraphics[width=1.0\linewidth]{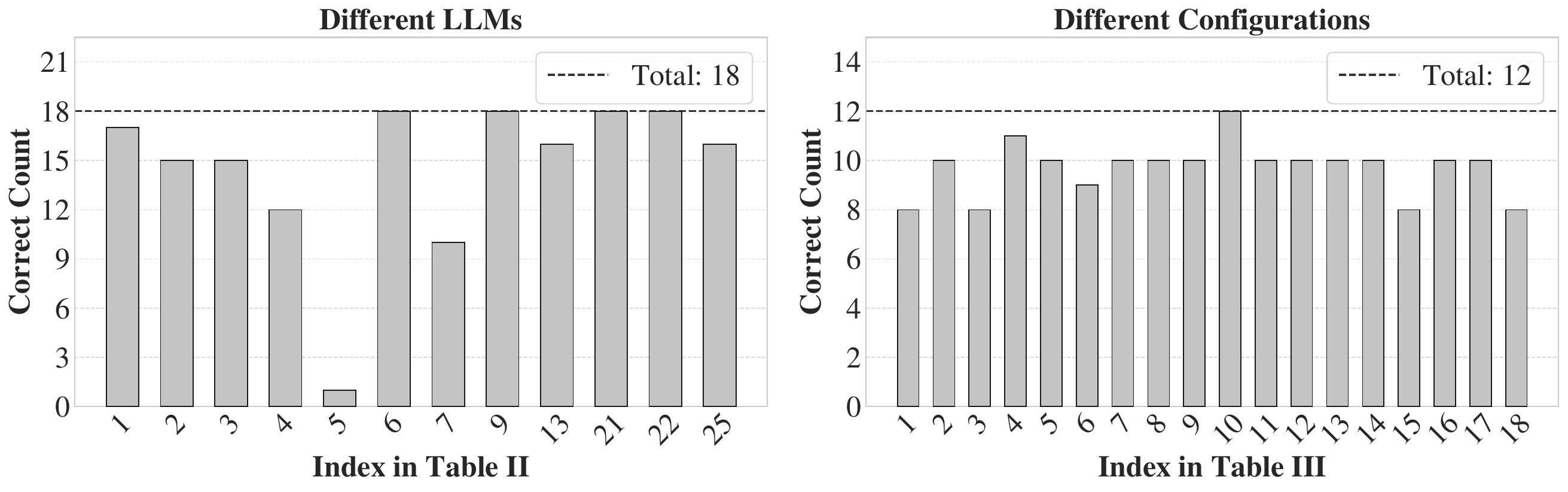}
    \caption{Top-1 accuracy across different LLMs and service configurations.}
    \label{fig:config_and_model}
\end{figure*}

\noindent\textbf{Fine-grained Analysis.} Figure~\ref{fig:config_and_model} presents the Top-1 accuracy of \toolname{} across different LLM service configurations and LLM identities. From the LLM perspective, all LLMs except \texttt{mimo-v2.5} correctly solve at least 10 out of 18 instances, and four LLMs achieve perfect identification accuracy, indicating that \toolname exhibits good generalizability across different LLMs. From the configuration perspective, all configurations correctly solve at least 8 out of 12 instances, indicating that \toolname exhibits good generalizability across different service configurations. We observe that \texttt{mimo-v2.5} failures are mainly caused by verbosity-induced mismatch. It often produces much longer responses, and the Jaccard similarity penalizes these extra n-grams through a larger
union, making \texttt{mimo-v2.5} appears less similar even when it covers the target content. To diagnose this issue, we replace Jaccard similarity with a target-side containment score:
\begin{equation}
\operatorname{Containment}_n(x,y)
=
\frac{|G_n(x)\cap G_n(y)|}{|G_n(y)|}.
\end{equation}
Here, \(x\) denotes the candidate response and \(y\) denotes the target response, so the denominator uses
\(G_n(y)\) to measure how much of the target-side content is covered by the candidate.
With this length-calibrated metric, \toolname successfully identifies 13 out of 18 cases in Figure~\ref{fig:config_and_model}, suggesting that
the original failures are largely due to length-sensitive similarity measurement rather than the core mechanism of \toolname. However, this metric is not equally suitable for all models. When applied uniformly across all models, the Top-1, Top-3, and Top-5 accuracies drop to only 32.4\%, 64.8\%, and 81.5\%, respectively. The containment metric is suitable for evaluating longer responses, but it is less broadly applicable than the Jaccard metric.
We also present the heatmaps of the average similarity rankings among all LLMs in Appendix~\ref{sec:heatmap}. For almost all target models, the top-1 ranked candidate corresponds to the same underlying entity as the target model.

\noindent\textbf{Resistance to Defense Mechanisms.} To assess the effectiveness of different methods in defense scenarios, we consider both input-side and output-side validation. Input-side validation checks whether the LLM fingerprinting queries are safe and topic-relevant, while output-side validation examines whether the responses generated by the target LLM are safe and policy-compliant. Such guardrails have been widely deployed in real-world LLM services~\cite{meta2025llamapromptguard2-86m,openai2025guardrails,meta2024llamaguard3hf}.
\begin{table}[t]
\centering
\caption{Input validation across different methods.}
\label{tab:input_valiadation}
\setlength{\tabcolsep}{6.0pt}
\begin{tabular}{lccccc}
\toprule
\textbf{Method} 
& \textbf{LLMmap} 
& \textbf{Met}
& \textbf{ZeroPrint}
& \textbf{\toolname}
\\
\midrule
\textbf{PromptGuard} & 3/8 & 0/25 & 0/10 & 0/48 \\
\textbf{Jailbreak} & 4/8 & 0/25 & 0/10 & 0/48 \\
\textbf{Off-Topic} & 133/144 & 399/450 & 112/180 & 3/54 \\
\bottomrule
\end{tabular}
\end{table}

For input-side validation, we employ Llama-Prompt-Guard-2-86M~\cite{meta2025llamapromptguard2-86m} and OpenAI Guardrails~\cite{openai2025guardrails}, two representative guardrail mechanisms used in LLM services. Prompt-Guard-2-86M (PromptGuard) evaluates prompt-injection risks. Guardrails\_Jailbreak (Jailbreak) detects jailbreak attempts, while
Guardrails\_Off\_Topic (Off-Topic) identifies off-topic queries. LLMmap, Met, and ZeroPrint include 8, 25, and 10 different default queries. For \toolname, we construct three groups of initial questions for each of the 16 different assistant domains listed in Table~\ref{tab:prompt_configs}, yielding 48 initial questions in total. For off-topic query detection, the initial questions must be dynamically adapted to the specific system prompt of each LLM service. The four methods contain 144, 450, 180, and 54 distinct initial questions, respectively. The detailed results are reported in Table~\ref{tab:input_valiadation}.
LLMmap relies on safety-alignment questions that often trigger jailbreak or prompt-injection rejections. Moreover, LLMmap, Met, and ZeroPrint use fixed domain-specific questions, resulting in off-topic rejection rates exceeding 92\%, 88\%, and 62\%, respectively. In contrast, only 3 of the 54 initial questions generated by \toolname are off-topic. Since \toolname does not depend on fixed query questions and incurs only a low cost for generating new questions, these results highlight its superiority in real-world LLM service scenarios.

For output-side validation, we integrate a harmful-content risk assessment module into the OpenAI Guardrails framework. About 1\% of \toolname's responses are flagged as posing potential safety risks. Manual analysis suggests that most flags arise from the medical and fitness domains of the target LLM services rather than explicitly harmful content.

\begin{figure}[t]
    \centering
    \includegraphics[width=1.0\linewidth]{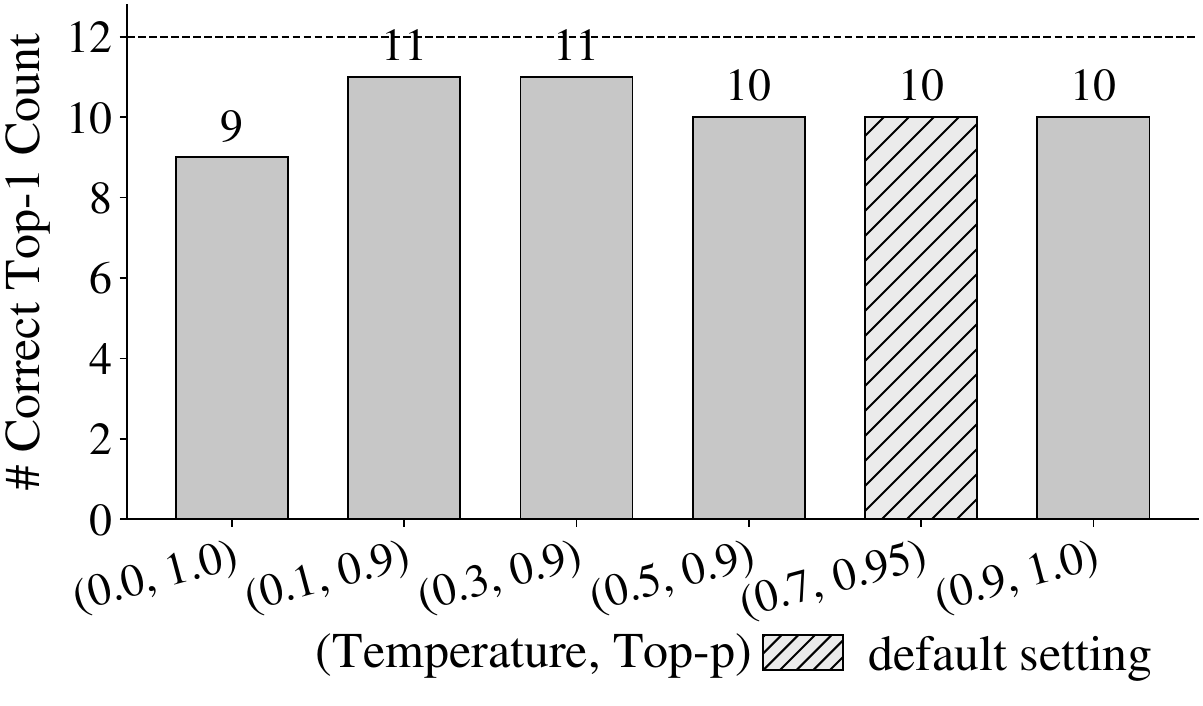}
    \caption{Impact of decoding settings on top-1 
    correct count.}
    \label{fig:different_decoing_settings}
\end{figure}

\noindent\textbf{Robustness to Decoding Randomness.} To evaluate the robustness of \toolname to variations in decoding parameters, we randomly select two assistants, No. 5 and No. 10 in Table~\ref{tab:prompt_configs}, assigning Models No. 1-6 in Table~\ref{tab:candidate_models} to the former and Models No. 7, 9, 13, 21, 22, and 25 to the latter. In addition to the original decoding setting of temperature and top-p, $(0.7, 0.95)$, we evaluate \toolname under five additional settings: $(0.0, 1.0)$, $(0.1, 0.9)$, $(0.3, 0.9)$, $(0.5, 0.9)$, and $(0.9, 1.0)$.
As shown in Figure~\ref{fig:different_decoing_settings}, \toolname maintains consistently effective performance across decoding settings with different levels of randomness.

\noindent\textbf{Open-Set Performance.} We primarily evaluate \toolname in the closed-set setting, solely to provide a more intuitive demonstration and comparison of its effectiveness. The effectiveness of \toolname does not rely on the closed-set assumption that the candidate model pool necessarily contains the target model. In the open-set scenario, \toolname is required to either make a definitive prediction of the model identity or determine that the target model is none of the candidates. A high absolute similarity threshold alone is not sufficient to support a reliable identification decision, because different target instances may have different score scales. Instead of applying a single global threshold to the raw similarity score, \toolname uses the score gap between the top-1 and top-2 candidates. Let \(\mathrm{sim}_{\mathrm{top1}}\) and \(\mathrm{sim}_{\mathrm{top2}}\) denote the highest and second-highest similarity scores between
\(\mathrm{LLM}_{\mathrm{black}}\) and the candidate models. \toolname accepts the top-1 prediction only when
\begin{equation}
\Delta_{\mathrm{top}}
=
\mathrm{sim}_{\mathrm{top1}}
-
\mathrm{sim}_{\mathrm{top2}}
\geq \tau .
\end{equation}
Otherwise, the prediction is rejected as ``none-of-all". Here, \(\tau>0\) requires the top-1 candidate to be sufficiently separated from the runner-up, instead of being accepted merely because it ranks first.

We evaluate open-set rejection under two candidate-pool settings: a 12-model candidate pool containing the target models and the full 27-model candidate pool. We consider both target-present and target-absent cases. In the target-present case, the ground-truth model is included in the candidate pool, and \toolname either accepts or rejects its top-1 prediction according to the rejection threshold $\tau$. In the target-absent case, the ground-truth model is removed from the candidate pool; therefore, any accepted top-1 prediction is necessarily a false acceptance of an unknown target.

For each threshold $\tau$, we report three counts. \textbf{Correct Known Accepts (CKA)} denotes the number of target-present cases for which the top-1 prediction is accepted and matches the ground-truth model. \textbf{Incorrect Known Accepts (IKA)} denotes the number of target-present cases for which the top-1 prediction is accepted but does not match the ground-truth model. \textbf{Unknown Accepts (UA)} denotes the number of target-absent cases for which \toolname incorrectly accepts a candidate model instead of rejecting as unknown. We further summarize these outcomes using three metrics. \textbf{Known Acceptance Precision (KAP)} measures the accuracy of accepted predictions among target-present cases and is defined as $\mathrm{KAP}=\mathrm{CKA}/(\mathrm{CKA}+\mathrm{IKA})$. \textbf{Known Correct Acceptance Rate (KCAR)} measures the fraction of all target-present cases that are both accepted and correctly identified, and is defined as $\mathrm{KCAR}=\mathrm{CKA}/N_{\mathrm{known}}$, where $N_{\mathrm{known}}$ is the total number of target-present cases. \textbf{Unknown Rejection Rate (URR)} measures the fraction of target-absent cases that are correctly rejected, and is defined as $\mathrm{URR}=1-\mathrm{UA}/N_{\mathrm{unknown}}$, where $N_{\mathrm{unknown}}$ is the total number of target-absent cases.

\begin{figure}[t]
    \centering
    \includegraphics[width=1.0\linewidth]{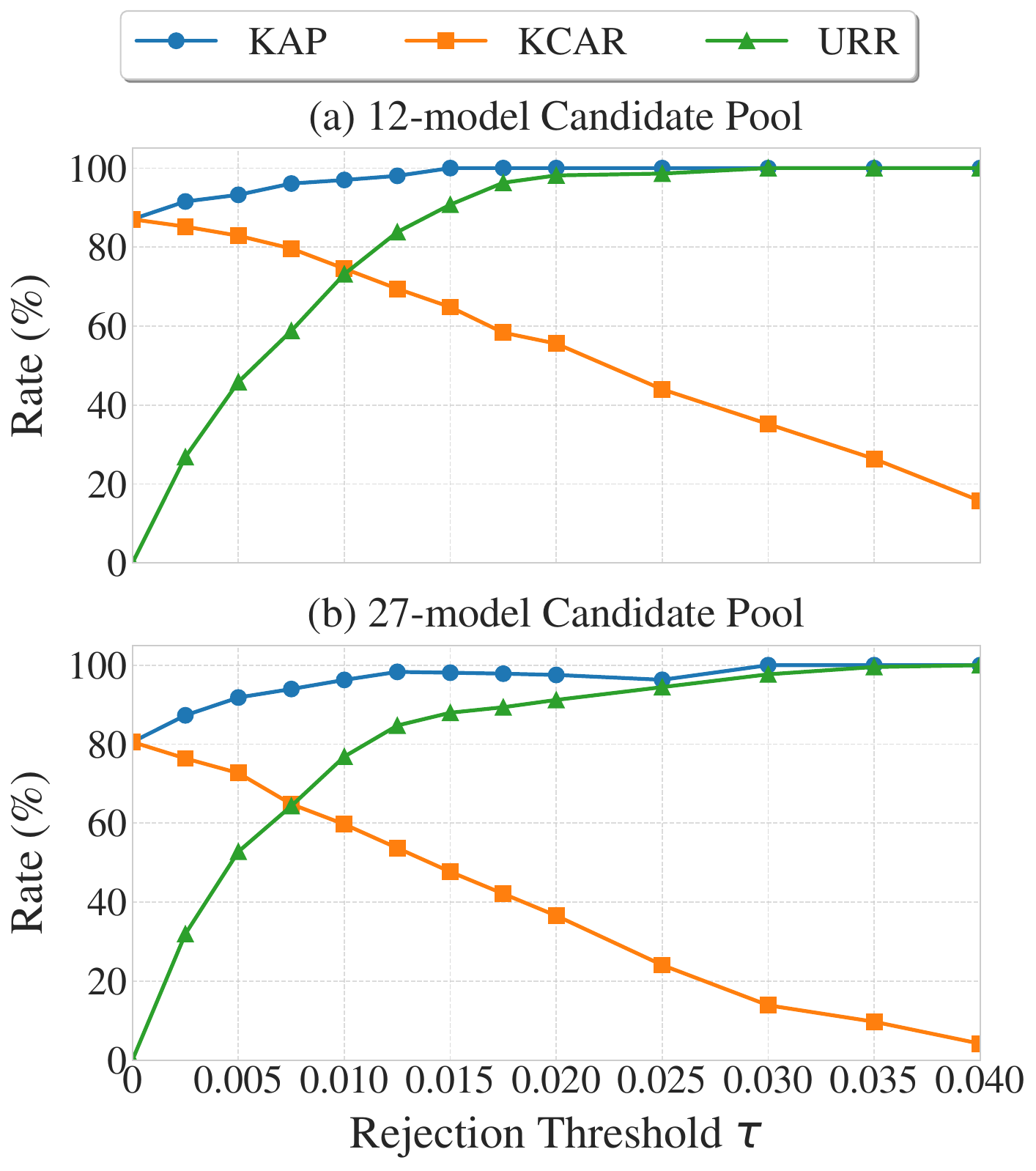}
    \caption{Open-set results under different margin thresholds.}
    \label{fig:openset_rejection}
\end{figure}

The results are shown in Figure~\ref{fig:openset_rejection}. Increasing the rejection threshold $\tau$ leads to a clear trade-off between known-target acceptance and unknown-target rejection. Specifically, KAP and URR generally increase with $\tau$, indicating that accepted predictions become more reliable and target-absent cases are more effectively rejected, whereas KCAR decreases as increasingly strict rejection removes more correct known-target predictions. Overall, $\tau=0.01$ and $\tau=0.0125$ achieves a good balance in both settings, maintaining high KAP while achieving a reasonable balance between KCAR and URR.
It is worth noting that the baseline methods are primarily designed for similarity-based matching, and they do not explicitly discuss threshold calibration for ``none of all" rejection. Therefore, their support for this scenario is limited.

\subsection{RQ2: Ablation Study}
\begin{table}[t]
\centering
\caption{Ablation study of probing strategies.}
\label{tab:ablation1}
\setlength{\tabcolsep}{6pt}
\begin{tabular}{lccc }
\toprule
\textbf{Strategy} 
& \textbf{Top-1} 
& \textbf{Top-3}
& \textbf{Top-5}
\\
\midrule
\textbf{Direct}
& 61.6\% & 79.6\% & 87.0\% \\ 

\textbf{Continuation} 
& 66.2\% & 81.9\% & 87.5\%  \\ 

\textbf{Follow-up} 
& 67.6\% & 85.7\% & 89.4\%  \\ 

\textbf{Direct \& Continuation}
& 73.6\% & 87.0\% & 89.8\%  \\ 

\textbf{Direct \& Follow-up}
& 74.5\% & 88.4\% & 91.2\%\\ 

\textbf{Follow-up \&  Continuation}
& 75.9\% & 89.8\% & 92.6\%\\ 

\textbf{Direct \&  Continuation  \& Follow-up}
& 80.6\% & 90.3\% & 92.1\% \\ 
\bottomrule
\end{tabular}
\end{table}

\begin{table}[t]
\centering
\caption{Ablation of embedding models and metrics.}
\label{tab:ablation2}
\setlength{\tabcolsep}{6.5pt}
\begin{tabular}{lccc}
\toprule
\textbf{Ablation Settings} 
& \textbf{Top-1} 
& \textbf{Top-3}
& \textbf{Top-5}
\\
\midrule
\textbf{base} & 56.0\% & 82.4\% & 88.4\% \\
\midrule
\textbf{multilingual-e5-large-instruct} & 59.7\% & 81.5\% & 90.7\% \\
\textbf{bge-large-en-v1.5} & 59.3\% & 81.0\% & 87.5\% \\
\textbf{edit distance} & 56.9\% & 74.5\% & 80.6\% \\
\textbf{longest common substring} & 54.2\% & 70.8\% & 76.9\% \\
\bottomrule
\end{tabular}
% }
\end{table}

\begin{figure}[t]
    \centering
    \includegraphics[width=1.0\columnwidth]{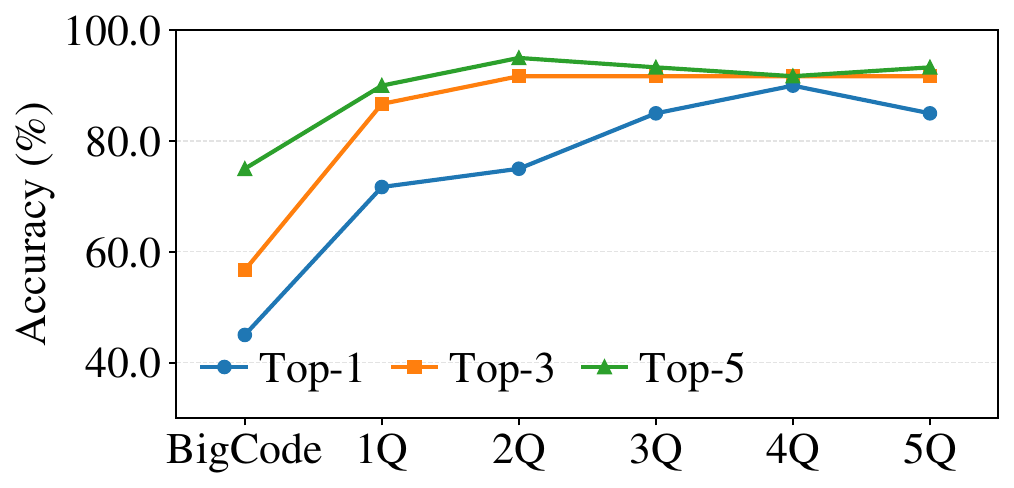}
    \caption{Accuracy across different numbers of questions.}
    \label{fig:trend}
\end{figure}

To answer RQ2, we analyze the impact of each component of \toolname{} on its effectiveness. 

\noindent\textbf{Ablation of Strategies. }
As shown in Table~\ref{tab:ablation1}, we decompose \toolname{}-Triple in Table~\ref{tab:main_experiment} by examining the performance of individual probing strategies and their pairwise combinations. The results show that these strategies provide complementary support and jointly enhance the effectiveness of \toolname{}. For the computation of semantic similarity and textual similarity, we conduct evaluations using different embedding models and textual similarity metrics, respectively. Specifically, we base on the results of \toolname{}-Q1 in Table~\ref{tab:main_experiment}. For semantic similarity, we additionally select two embedding models: \texttt{multilingual-e5-large-instruct}~\cite{wang2024multilingual} and \texttt{bge-large-en-v1.5}~\cite{baai2023bgelargeenv15}. For textual similarity, we adopt edit distance and longest common substring as the evaluation metrics. The results are shown in Table~\ref{tab:ablation2}. Different embedding models achieve comparable performance, as they mainly serve to extract semantic representations. Jaccard similarity achieves better performance in textual similarity measurement because it focuses on the overlap of local textual units rather than strict sequence-level alignment. In contrast, edit distance and longest common substring metrics are more sensitive to sequential changes and length differences, and may lead to inaccurate similarity judgments for texts that differ in surface.

\noindent\textbf{Impact of Questions and Continuation Starting Points. } The performance of \toolname{} improves as the number of initial questions increases. We further analyze this situation in depth. Specifically, based on the results of \toolname{}-Q1 in Table~\ref{tab:main_experiment}, we select the subsets corresponding to No. 1, No. 4, No. 5, No. 7, and No. 10 in Table~\ref{tab:prompt_configs}, which cover a range of Top-1 accuracy from low to high. To further evaluate the quality of the initial question, we randomly choose a question from BigCodebench~\cite{zhuo2024bigcodebench}.
The experimental results are shown in Figure~\ref{fig:trend}. The accuracy of \toolname{} improves steadily as the number of questions increases. Using multiple questions helps reduce the randomness and noise in LLM responses and enables the extraction of more stable model fingerprints. In comparison, the initial question sampled from BigCodebench results in substantially lower performance, demonstrating the effectiveness of adaptive initial question generation. To further validate the effectiveness of the continuation starting point selection in Algorithm 1, we replace it with three randomly selected starting points under each of the five configurations. Compared with \toolname{}-Q1 when considering the continuation strategy in isolation, the Top-1, Top-3, and Top-5 accuracies decrease from 53.3\%, 73.3\%, and 83.3\% to 48.3\%, 66.7\%, and 71.7\%, respectively, demonstrating the benefit of the proposed selection strategy.

\begin{figure}[t]
    \centering
    \includegraphics[width=1.0\linewidth]{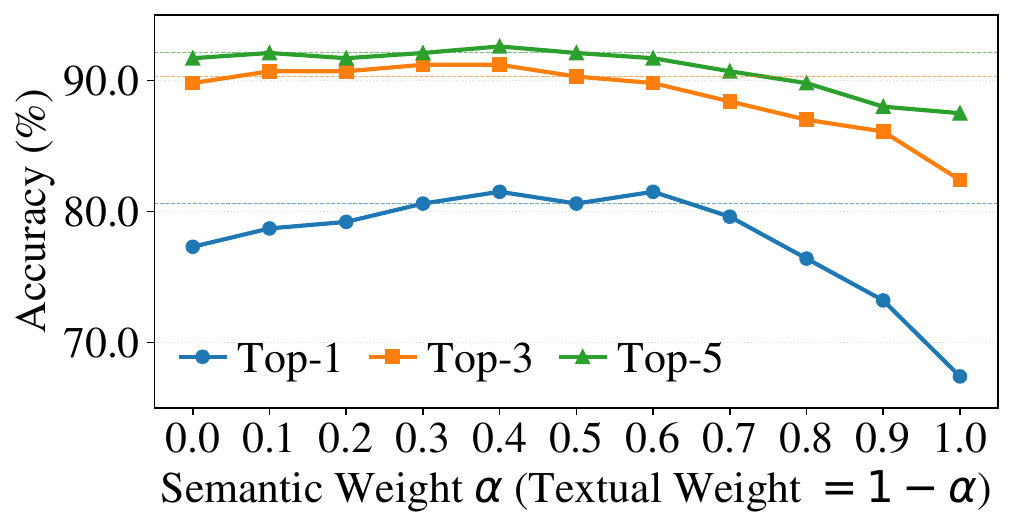}
    \caption{Weight ablation of \toolname{}.}
    \label{fig:weight_ablation}
\end{figure}

\noindent\textbf{Impact of weighting ratios. }We also evaluate different weighting ratios between semantic and textual similarities, as shown in Figure~\ref{fig:weight_ablation}. Relying solely on either similarity measure results in relatively lower performance, whereas their combination consistently achieves competitive results. Accordingly, we adopt equal weights of 0.5 as the default setting. This choice is not intended to be optimal, but provides a general and effective setting without additional parameter tuning.

\subsection{RQ3: Applications}
To answer RQ3, we conduct a case study on real-world LLM service and further evaluate the performance of \toolname{} on fine-tuned LLMs.

\noindent\textbf{Case Study.} We demonstrate the workflow of \toolname on a Python Expert Chatbot hosted on POE~\cite{nihiniverse2026pythonexpert4o}. This chatbot publicly discloses its underlying model as \texttt{gpt-4o-mini}, while other configuration details remain black-box. We select this target for two reasons. First, the disclosed model identity allows us to obtain a reliable ground truth while still emulating the LLM configuration uncertainty of a black-box LLM service. Second, the ultimate goal of \toolname is to infer the backend LLM identity of a service through benign user interactions. In this setting, where the LLM identity is already publicly available, our analysis does not compromise any rights or interests of the service provider.
\begin{figure}
    \centering
    \includegraphics[width=1.0\linewidth]{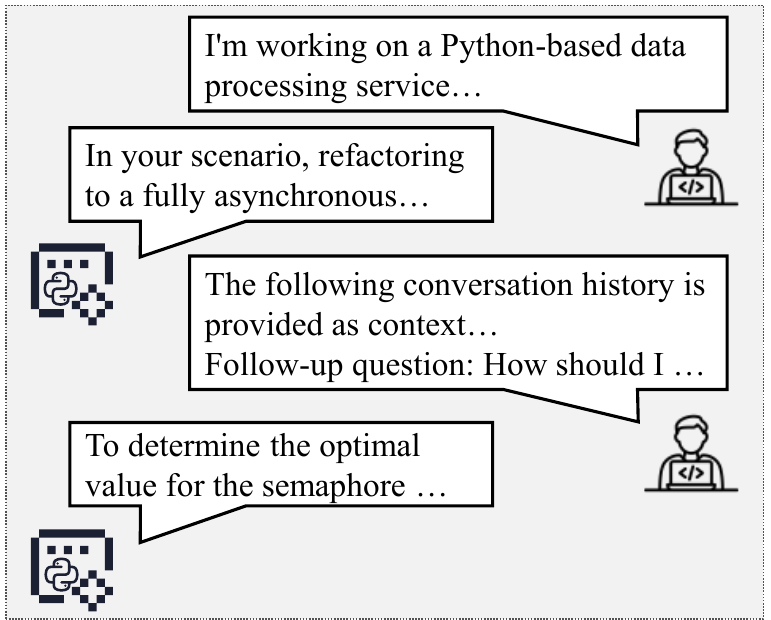}
    \caption{Case study of \toolname{}.}
    \label{fig:case_study}
\end{figure}
Figure~\ref{fig:case_study} illustrates the interaction process. \toolname issues two queries to the LLM service: the first is used for direct probing and continuation probing, whereas the second is used for follow-up probing. Based on these interactions, \toolname successfully identifies the target model as \texttt{gpt-4o-mini}.

\noindent\textbf{Fine-tuned Models. }We select two LoRA-based fine-tuned models derived from \texttt{Llama-3.1-8B-Instruct} as the target LLMs: llama3.1-8b-reasoning-code math~\cite{yaxin1992_llama31_8b_reasoning_code_math} and flowertune-medical-lora-llama-3.1-8b-instruct~\cite{zjudai_flowertune_medical_lora_llama31_8b_instruct}, paired with system prompts No. 2 and No. 8 in Table~\ref{tab:prompt_configs}, respectively. The LLMs listed in Table~\ref{tab:candidate_models} are used as the candidate LLMs.
For each fine-tuned model, we evaluate \toolname under six temperature/top-p settings: 
$(0.0, 1.0)$, $(0.1, 0.9)$, $(0.3, 0.9)$, $(0.5, 0.9)$, $(0.7, 0.95)$, and $(0.9, 1.0)$. For these 12 fine-tuned settings, \toolname{} correctly identifies 8, 9, and 11 models at Top-1, Top-3, and Top-5, respectively. The results demonstrate that \toolname can effectively trace fine-tuned models back to their source models.

%% file: sections/discussion.tex
\section{Discussion}

\noindent\textbf{Limitations and Future Directions.}
The effectiveness of \toolname can be affected by probe question quality and dynamic multi-model routing. In our experiments, different probe questions may lead to different identification results even for the same LLM service under the same setting, highlighting the importance of generating informative and discriminative probes. Future work could explore more systematic probe optimization, such as reinforcement learning for targeted probe generation. In addition, \toolname assumes a stable underlying model within each probing session, which may not hold for services that dynamically route queries to different sub-LLMs. Such routing can reduce fingerprinting reliability. Since each response probing in \toolname uses only two consecutive, topic-consistent queries, the two queries are more likely to be routed to the same sub-LLM when routing is primarily determined by query semantics or task type. A systematic study of probe optimization and dynamic routing is left for future work.

\noindent\textbf{Cost Overhead. }Each initial probe question of 
\toolname requires querying the target model twice and querying each candidate model five times. Based on the three groups of initial probe questions used in Table~\ref{tab:main_experiment}, we further estimate the query cost, which averages only about \$0.044 per target black-box model. The cost is relatively low and can be reduced if open-source candidate models are deployed locally.

\noindent\textbf{Downsteam Tasks. }
\textit{(1) Security risk assessment of black-box LLM services.} Recent studies reveal that LLMs exhibit heterogeneous security properties across different domains and attack types~\cite{li2024salad,deepteam2025jailbreaking}, identifying the specific LLM behind a black-box service may enable attackers to exploit known model-specific vulnerabilities and launch more targeted attacks. For instance, knowledge of the underlying model can facilitate adaptive optimization of jailbreak prompts, achieving substantially higher attack success rates~\cite{andriushchenko2025jailbreaking,geisler2025reinforce}. The design of \toolname therefore helps assess the real-world security risks associated with model identity exposure and can support more effective security evaluation and hardening of black-box LLM services.
\textit{(2) Model copyright auditing and IP protection.} For model owners, \toolname can audit black-box LLM services for model provenance, helping detect unauthorized use or license violations and thereby supporting model copyright protection. It can also be extended to white-box settings to verify whether service providers faithfully deliver the claimed models through model similarity comparison, safeguarding subscribers' interests. By revealing the identity of the underlying LLM, \toolname further enables model owners to enforce usage policies and protect their intellectual property rights.

%% file: sections/related-work.tex
\section{Related Work}
\subsection{LLM Application Security}
As LLMs have been widely employed in various applications~\cite{topsakal2023creating, zhao2025llm, wu2024autogen}, they also raise security concerns due to their inherent susceptibility to jailbreaking, hallucinations, and prompt injection~\cite{zou2023universal,ren2024codeattack, perezignore, yu2025mind}. A malicious query perturbed by several exclamation marks, or embedded within a code snippet, can effectively bypass the guardrails of an aligned LLM~\cite{perezignore, andriushchenko2025jailbreaking}. More sophisticated attacks that leverage the internal states of the LLMs during response, or adaptively modify the malicious prompts based on the model's response, can further increase the attack success rates to up to 100\% ~\cite{liu2024making,yu2025mind, zhang2025bleeding, geisler2025reinforce}. The vulnerabilities of LLMs to attacks such as jailbreak and prompt injection not only reduce the reliability and validity of their responses but also introduce broader attack vectors for LLM-integrated applications. Configurations of LLMs, such as RAG datastores or system prompts, can be poisoned or embedded with backdoor triggers, enabling attacks targeting safety-critical applications \cite{ben2025gasliteing, chen2024agentpoison, zou2025poisonedrag}. Successful prompt injection attacks also fuel traditional attacks, including SQL injections and Remote Code Execution (RCE), towards LLM applications \cite{pedro2023prompt, liu2024demystifying}.

\subsection{LLM fingerprinting}
To protect LLM's intellectual property, white-box fingerprinting extracts distinctive features from model parameters and architecture. 
Xu et al. \cite{xu2024instructional} utilized instruction tuning to embed backdoor triggers in LLMs, which produce specific fingerprint texts upon activation.
Zeng et al. \cite{zeng2024huref} proposed HuRef, which identifies training-stable invariant terms in Transformers and is robust to weight rearrangement. 
Zhang et al. \cite{zhang2024reef} introduced REEF, a training-free method that compares the representations of LLMs using Centered Kernel Alignment (CKA) similarity. 
Wu et al. introduced TensorGuard, which fingerprints LLMs using statistical features of their gradients~\cite{wu2025tensorguard}.
% White-box LLM fingerprinting relies on fine-tuning LLMs, accessing LLMs' internal states, or the weights of LLMs. 
In contrast, black-box fingerprinting methods construct fingerprints by sending carefully designed queries to the target model and analyzing its output responses, since model parameters or internal representations are not available. 
% This calls for effective black-box fingerprinting techniques, and many studies have explored various approaches, as we show below. 
Pasquini et al. \cite{pasquini2025llmmap} proposed LLMmap, which uses greedy search to select queries that maximize inter-LLM differences while preserving intra-LLM consistency.
Gao et al. \cite{gao2024model} introduced MET, a statistical framework for Model Equality Testing (MET) that calculates Maximum Mean Discrepancy (MMD).
Shao et al.\cite{shao2026reading} designed ZeroPrint, which estimates LLM Jacobian-based fingerprints from query-response pairs and measures their similarity using Pearson correlation.

%% file: sections/conclusion.tex
\section{Conclusion}

In this paper, we focus on the problem of black-box LLM fingerprinting. We proposed \toolname{}, a response-adaptive fingerprinting method for revealing hidden LLM identities in black-box LLM services. 
\toolname{} integrates three response consistency probing strategies: Direct Probing, Continuation Probing, and Follow-up Probing. Extensive experimental results demonstrate that \toolname significantly outperforms SOTA methods, achieves strong robustness under different defense strategies and decoding parameters, and shows promising feasibility for practical applications.

%% file: sections/appendix.tex
\appendices

\subsection{Instrumental Prompts}
\label{sec:prompts}
\subsubsection{Initial Question Generation Prompt}
\label{sec:initial}
Here is the complete prompt for initial question generation.

\begin{shaded*}
\noindent
You are a realistic user-query generator for model fingerprinting. Generate one natural user request for the given task domain.\\

\noindent
The request should sound like something a real user would ask an AI assistant in that domain. It should be concrete enough to elicit a substantive answer, but it should not be over-specified, artificially technical, or framed like an evaluation benchmark.\\

\noindent
Requirements:\\
1) Use a realistic scenario with 3-5 concrete contextual details. At least one detail should introduce a constraint, failed prior attempt, conflicting preference, edge case, or risk boundary that makes the best answer non-obvious.\\
2) Make the request self-contained enough that the assistant can answer without asking follow-up questions.\\
3) The request should ask for practical judgment: prioritizing between options, explaining a tradeoff, setting a risk boundary, choosing a next step, or ruling out a tempting but unsuitable approach.\\
4) Avoid canonical FAQ-style questions or requests where most competent assistants would likely give the same common 2-3 suggestions.\\
5) Avoid asking for model identity, probabilities, token behavior, internal policies, or anything that looks like fingerprint testing.\\
6) Avoid rigid formats such as mandatory numbered lists, exhaustive checklists, or step-by-step derivations unless they are natural for the task.\\
7) Do not make the question intentionally obscure, adversarial, trick-like, or overly academic.\\
8) The answer should be comparable across models, but not forced into a single deterministic wording or structure.\\

\noindent
Given task: {task}\\

\noindent
Return only the final user question.
\end{shaded*}
% \end{tcolorbox}

\subsubsection{Continuing Prompt}
\label{sec:continuing}
Here is the complete continuing prompt.

\begin{shaded*}
\noindent
Here is the answer you generated last time, but it was cut off due to network issues.\\
Please recall it and continue the answer in the same context as before.\\

\noindent
Hard rules:\\
1) Output ONLY the continuation (no preface, no quotes, no JSON, no markdown).\\
2) NEVER ask for clarification.\\
3) NEVER mention that the input is incomplete/corrupted/truncated.\\
4) NEVER apologize or explain what you are doing.\\
5) NEVER repeat any part of the provided answer.\\
6) Preserve the same language, tone, and formatting style as the provided answer.\\
7) If the answer ends mid-word, first complete that word with the most likely characters, then continue.\\
8) Even if the provided answer is extremely short or looks fragmentary, you MUST still continue coherently.\\

\noindent
Question: \{question\}\\

\noindent
Previous\_answer: \{previous\_text\}\\

\noindent
Continue the answer from here: 
\end{shaded*}
% \end{tcolorbox}

\subsubsection{Following-up Question Generation Prompt}
\label{sec:following}
Here is the complete following-up question generation prompt.

\begin{shaded*}
\noindent
You are a conversation continuity assistant. Generate one natural follow-up question based on the given answer text.\\

\noindent
The question should sound like something a real user would ask after reading the answer. It should continue the same task or scenario, refer to a concrete point from the answer, and be specific enough to elicit a substantive response.\\

\noindent
Requirements:\\
1) Continue the same conversation without introducing an unrelated topic.\\
2) Refer to a specific decision, step, constraint, risk, example, or recommendation mentioned in the answer when possible.\\
3) Ask for a practical clarification, next action, tradeoff, edge case, or application detail.\\
4) Keep the question natural and concise; do not make it artificially technical, adversarial, or benchmark-like.\\
5) Do not ask about model identity, probabilities, tokens, internal policies, or hidden reasoning.\\
6) Avoid generic follow-ups such as "Can you elaborate?" or "Tell me more?"\\
7) Do not force a rigid format such as a numbered list or step-by-step derivation unless that is natural for the conversation.\\

\noindent
Output rules:\\
- Output exactly one question.\\
- Do not include explanations, prefixes, or quotation marks.\\

\noindent
Text: \{text\}\\
\end{shaded*}
% \end{tcolorbox}

\subsection{CoT and RAG Prompt Templates}
Here are the prompt templates of Chain-of-Thought and Retrieval-Augmented Generation.

\begin{shaded*}
    
\noindent
\textbf{Fitness Tracker: }\{user\_query\} + ``Let's think step by step."\\
\textbf{Travel Planning: }``Please provide a step-by-step explanation and a final answer: " + \{user\_query\}\\
\textbf{Tutoring Assistance: }``Let's think step by step and then give the final answer: " + \{user\_query\}\\
\textbf{Coding Troubleshooting \& Medical Diagnosis II: }Context Details:\\\{context\}\\ \\
Question:\\\{question\}\\ \\
Answer based on the provided context: 
\end{shaded*}
% \end{tcolorbox}

\subsection{Generic System Prompt}
\label{sec:generic}
Here is the generic system prompt we used.
\begin{shaded*}

\noindent
You are a helpful, honest, and efficient AI assistant.\\
Answer clearly, accurately, and concisely. \\
Understand the user's intent, give practical and actionable help, and explain complex topics step by step when needed.
    
\end{shaded*}
% \end{tcolorbox}

\subsection{Examples}
\subsubsection{Initial Question Examples}
Here are some initial question examples.

\begin{shaded*}
\noindent
\textbf{Example I (shopping): }I'm looking to buy a high-end espresso machine for my kitchen, and I've narrowed it down to the Decent DE1Pro and the La Marzocco Linea Micra. I drink mostly medium-dark roast traditional lattes, but my partner is really into experimenting with light roast ``turbo shots" and terroir exploration. We have limited counter space under some low-hanging cabinets, and while I’m willing to spend the \$3,000+, I’m concerned because I’ve heard the Decent can feel a bit like operating a computer rather than a tactile appliance, whereas the Micra might be too ``locked-in" for the experimental shots my partner wants. Given that we need something that heats up in under 10 minutes for our morning commute but won't become an obsolete piece of hardware in five years, which of these two should I prioritize, or is there a specific tradeoff in the workflow of one that makes it a dealbreaker for this specific mix of users?

\noindent
\textbf{Example II (pet care): }My 8-year-old indoor cat, Jasper, has suddenly started urinating on the bathmat and my laundry pile over the last three days, even though his litter box is kept clean and hasn't changed location. He seems to be acting normally otherwise—eating well and still playful—but I did recently start a new job where I'm gone for 10 hours a day instead of working from home. I'm torn between taking him to the emergency vet tonight because I'm worried about a blockage, or waiting a few days to see if he's just stressed and acting out because of my new schedule. Given that he is still successfully passing urine (just in the wrong places) and isn't straining or crying, what is the most responsible way to prioritize his care right now without causing him unnecessary stress or deanonymizing my bank account on an emergency visit?
\end{shaded*}
% \end{tcolorbox}

\subsection{Candidate Model Rank}
\label{sec:heatmap}
\begin{figure*}[t]
    \centering
    \includegraphics[width=1.0\linewidth]{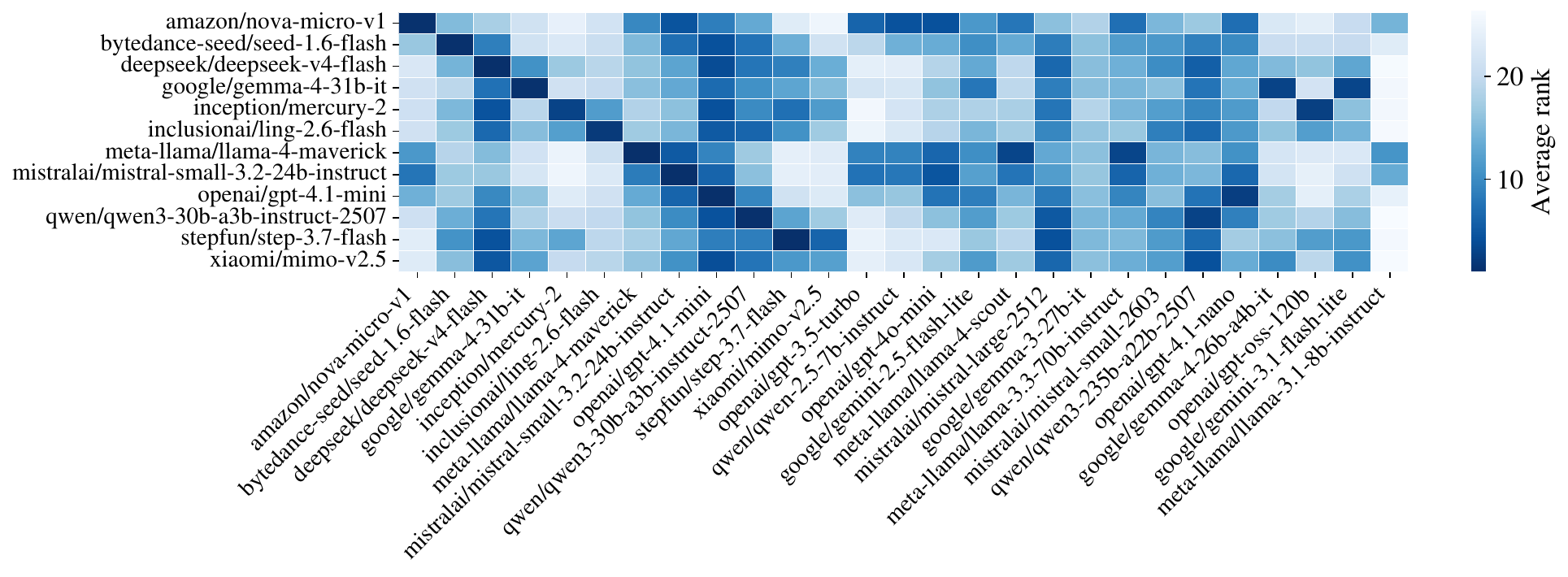}
    \caption{Average Candidate Rank by Target Model.}
    \label{fig:heatmap}
\end{figure*}
The average candidate rank by target model is shown in Figure~\ref{fig:heatmap}.